\documentclass[twocolumn,english,aps,prl,showpacs]{revtex4}
\usepackage[T1]{fontenc}
\usepackage[latin1]{inputenc}
\usepackage{amsmath}
\usepackage{graphicx}
\usepackage{amssymb}
\usepackage{esint}

\makeatletter
\@ifundefined{textcolor}{}
{%
 \definecolor{BLACK}{gray}{0}
 \definecolor{WHITE}{gray}{1}
 \definecolor{RED}{rgb}{1,0,0}
 \definecolor{GREEN}{rgb}{0,1,0}
 \definecolor{BLUE}{rgb}{0,0,1}
 \definecolor{CYAN}{cmyk}{1,0,0,0}
 \definecolor{MAGENTA}{cmyk}{0,1,0,0}
 \definecolor{YELLOW}{cmyk}{0,0,1,0}
 }

\makeatother

\makeatother

\usepackage{babel}

\makeatother

\usepackage{babel}

\makeatother

\usepackage{babel}

\makeatother

\usepackage{babel}

\begin{document}

\title{Pseudo-gap pairing in ultracold Fermi atoms}

\author{Hui Hu$^{1}$, Xia-Ji Liu$^{1}$, Peter D. Drummond$^{1}$, }

\thanks{E-mail: pdrummond@swin.edu.au}

\author{Hui Dong$^{1,2}$}

\affiliation{$^{1}$\ ARC Centre of Excellence for Quantum-Atom Optics and Centre
for Atom Optics and Ultrafast Spectroscopy, Swinburne University of
Technology, Melbourne 3122, Australia \\
 $^{2}$\ Institute of Theoretical Physics, The Chinese Academy
of Sciences, Beijing 100080, China}

\date{\today{}}
\begin{abstract}
\noindent The BEC-BCS crossover in ultracold Fermi gases creates an
ideal environment to enrich our knowledge of many-body systems. It
is relevant to a wide range of fields from condensed matter to astrophysics.
The nature of pairing in strongly interacting Fermi gases can be readily
studied. This aids our understanding of related problems in high-$T_{c}$
superconductors, whose mechanism is still under debate due to the
large interaction parameter. Here, we calculate the dynamical properties
of a normal, trapped strongly correlated Fermi gas, by developing
a quantum cluster expansion. Our calculations for the single-particle
spectral function agree with recent rf spectroscopy measurements,
and clearly demonstrate pseudogap pairing in the strongly interacting
regime. 
\end{abstract}


\maketitle
Ultracold Fermi atom experiments allow inter-particle interactions,
geometries, and spin species of atomic gases to be precisely controlled
and tuned at will, leading to an exactly known model Hamiltonian \cite{giorgini}.
Since rapid experimental progress is able to provide accurate data
on both static \cite{luo,horikoshi,nascimbene} and dynamic \cite{chin,stewart,veeravalli,gaebler}
properties, this type of \emph{Quantum Simulation} can definitively
settle fundamental issues. An interesting example is momentum-resolved
rf spectroscopy at the BEC-BCS crossover \cite{stewart,gaebler},
which can reveal single-particle excitation gaps or pseudogaps in
the normal state, as in analogous high-$T_{c}$ superconductor systems
\cite{randeria1992,randeria1995,damascelli,kanigel,lehur}.

By contrast, theoretical progress \cite{leggett,nsr,randeria1993,ohashi,hldepl}
in describing strong pairing fluctuations at the BEC-BCS crossover
is notoriously difficult due to the lack of a small interaction parameter
\cite{hldnjp}. The situation is most severe for dynamical properties,
where different crossover theories lead to qualitatively different
predictions. For the single-particle spectral function in the strong-coupling
regime, some crossover theories predict a pseudogap - the precursor
of fermionic pairing in the normal state above $T_{c}$ \cite{perali,bruun2006,bruun2008,tsuchiya,chen}
- while some others \cite{haussmann} claim no such effects. Quantum
Monte Carlo simulation of the spectral function is not conclusive
\cite{magierski}.

In this Letter, we develop a quantum virial or cluster expansion to
solve this delicate problem of dynamic properties for a normal, trapped,
and strongly interacting Fermi gas. The advantages of this method
are clear. First, the expansion has a controllable parameter \cite{hove,liuve}:
the fugacity $z=\exp(\mu/k_{B}T)$ is small at high temperatures $T$.
Second, the expansion coefficient or function at the $n$-th order
($n\geqslant2$) is entirely determined by knowledge of a $n$-particle
cluster. Multiparticle correlations, which are missing in most current
crossover theories, can be easily accounted for and improved. Finally,
the method is easily capable of treating external potentials or traps.
We have estimated the validity of the expansion by comparing its predictions
with experimental data for thermodynamics \cite{hldnjp}. At the cusp
of the crossover, where the \textit{s}-wave scattering length $a_{s}$
diverges (unitarity limit \cite{hldnatphys}), it is applicable down
to $0.4T_{F}$ for a trapped Fermi gas \cite{hldnjp}.

To develop the cluster expansion for general dynamical properties,
let us consider two arbitrary linear operators of physical interest,
$\hat{R}$ and $\hat{S}$, and the related Green function or correlation
function at different space-time points,\begin{equation}
G\left({\bf r},{\bf r}^{\prime};\tau\right)\equiv-\frac{\text{Tr}\left[e^{-\beta\left({\cal H}-\mu{\cal N}\right)}\hat{R}\left({\bf r,}\tau\right)\hat{S}^{+}\left({\bf r}^{\prime}\right)\right]}{\text{Tr}e^{-\beta\left({\cal H}-\mu{\cal N}\right)}},\end{equation}
where at finite temperatures we are working with an imaginary time
$\tau$ in the interval $0<\tau\leq\beta=1/k_{B}T$. At high temperatures,
both numerator and denominator may be expanded into the powers of
$z\ll1$, leading to $G\left({\bf r},{\bf r}^{\prime};\tau\right)=(X_{0}+zX_{1}+\cdots)/(1+zQ_{1}+\cdots)=X_{0}+z\left(X_{1}-X_{0}Q_{1}\right)+\cdots$,
where $X_{n}=-$ Tr$_{n}[e^{-\beta{\cal H}}\hat{R}\left({\bf r,}\tau\right)\hat{S}^{+}({\bf r}^{\prime})]$
is the expansion function and $Q_{n}=$Tr$_{n}[e^{-\beta{\cal H}}]$
is the cluster partition function. The above expansion is to be referred
to as the cluster expansion of correlation function, $G\left({\bf r},{\bf r}^{\prime};\tau\right)=G^{(0)}\left({\bf r},{\bf r}^{\prime};\tau\right)+zG^{(1)}\left({\bf r},{\bf r}^{\prime};\tau\right)+\cdots,$
where, \begin{eqnarray}
G^{(0)}\left({\bf r},{\bf r}^{\prime};\tau\right) & = & X_{0},\nonumber \\
G^{(1)}\left({\bf r},{\bf r}^{\prime};\tau\right) & = & X_{1}-X_{0}Q_{1},\ \text{etc}.\end{eqnarray}
The experimentally measured spectral function $A\left({\bf k},\omega\right)$
can be calculated from the correlation function via analytic continuation,
so that we may write accordingly, $A\left({\bf k},\omega\right)=A^{(0)}\left({\bf k},\omega\right)+zA^{(1)}\left({\bf k},\omega\right)+\cdots$.
The calculation of the $n$-th expansion coefficient $G^{(n)}\left({\bf r},{\bf r}^{\prime};\tau\right)$
or $A^{(n)}\left({\bf k},\omega\right)$ requires the knowledge of
solutions up to the $n$-body problem, including both energy levels
and wavefunctions (see Appendix). In this work we aim to calculate
the leading effects, which give the 2$^{\textrm{nd}}$-order expansion
function. The next-order expansion function \cite{liuve}, is not
treated here for simplicity.

\begin{figure}[htp]

\begin{centering}
\includegraphics[clip,width=0.45\textwidth]{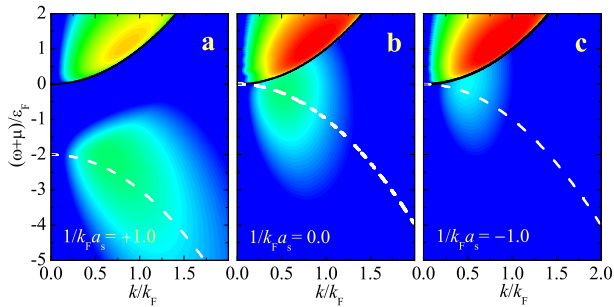} 
\par\end{centering}

\caption{(Color online) Contour plots of the occupied spectral intensity at
crossover. The intensity $I(\omega)=A({\bf k},\omega)f(\omega)k^{2}/(2\pi^{2})$
increases from blue ($10^{-3}I_{max}$) to red ($I_{max}$) in a logarithmic
scale. The calculations were performed with harmonic traps at $T=0.7T_{F}$
and $1/(k_{F}a_{s})=+1$, $0$, $-1$, with a resulting fugacity at
the trap center of $z\simeq0.14$, $0.42$, and $0.48$, respectively.}

\label{fig1} 
\end{figure}

To calculate the single-particle spectral function $A({\bf k},\omega)$,
we take the annihilation field operators $\hat{\Psi}_{\sigma}(\mathbf{r},\tau)$
($\sigma=\uparrow,\downarrow$) for $\hat{R}$ or $\hat{S}$ (see
Appendix). Fig. 1 shows contour plots of the \emph{occupied} spectral
intensity of a trapped Fermi gas in the crossover at $T=0.7T_{F}$.
At this temperature, our results are \emph{quantitatively} reliable
\cite{hldnjp}. We observe that, in addition to the response from
coherent Landau quasiparticles (black lines), there is a broad incoherent
spectral weight centered about $\omega+\mu=-\epsilon_{\mathbf{k}}-\epsilon_{B}$
(white dashed lines), where $\epsilon_{\mathbf{k}}=\hbar^{2}k^{2}/(2m)$
and $\epsilon_{B}=\hbar^{2}/(ma_{s}^{2})$ is the binding energy.
Thus, the spectra clearly exhibit a gap-like double peak structure
in the normal state. This is a remarkable feature: the dispersion
at negative energies seems to follow the BCS-like dispersion curve,
$\omega=-\sqrt{(\epsilon_{\mathbf{k}}-\mu)^{2}+\triangle^{2}}$, and
behaves as if the gas was superconducting, even though we are above
the critical temperature $T_{c}$. Therefore, the incoherent spectral
weight indicates the tendency of pseudogap: the precursor of fermionic
pairing due to strong attractions, i.e., it arises from the atoms
in the paired state or {}``molecules''. The pairing response is
very broad in energy and bends down towards lower energy for increasing
$k$. At large $1/(k_{F}a_{s})$, the width is of order $\sqrt{\mbox{max}\{k_{B}T,\epsilon_{F}\}\epsilon_{\mathbf{k}}}$.

The incoherent spectral weight found by our leading cluster expansion
is a universal feature of interacting Fermi gases. At large momentum
$k\gg k_{F}$, it is related to the universal $1/k^{4}$ tail of momentum
distribution \cite{schneider,tan}, $n_{\sigma}({\bf k})=\int_{-\infty}^{+\infty}d\omega A({\bf k},\omega)f(\omega)$$ $$\simeq\mathcal{{\normalcolor I}}/k^{4}$,
arising from the short-range two-body correlations. A nonzero contact
$\mathcal{I}$ therefore necessarily implies a finite spectral weight
at negative energies. At $k\sim k_{F}$, we anticipate that many-body
correlations will become increasingly important.

\begin{figure}[htp]
\begin{centering}
\includegraphics[clip,width=0.4\textwidth]{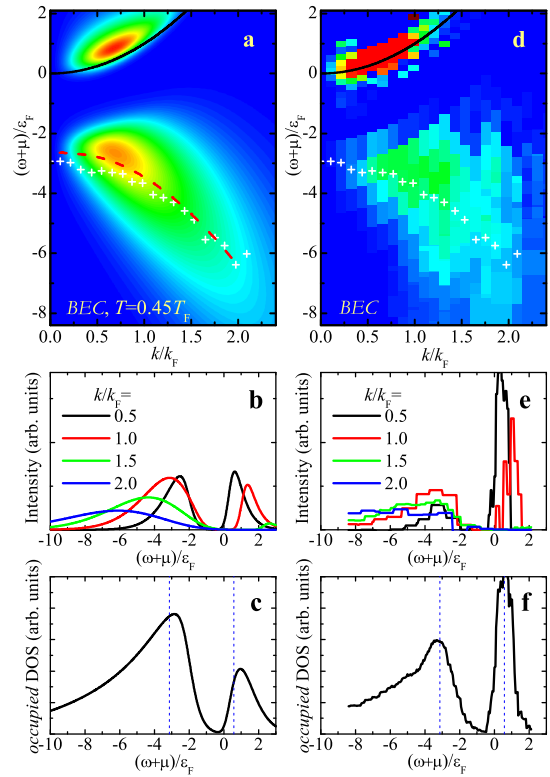} 
\par\end{centering}

\caption{(Color online) Single-particle excitation spectra on the BEC side
of crossover. \textbf{a-c}, Cluster expansion predictions ($z\simeq0.1$
and $\mu\simeq-1.08\epsilon_{F}$). \textbf{d-e}, Corresponding experimental
data \cite{stewart}. \textbf{a}, The linear-scale intensity map.
Our results were convoluted with a Gaussian broadening curve of width
$\sigma=0.22\epsilon_{F}$, to account for the measurement resolution
\cite{stewart}. The black line shows upper free-atom dispersion.
The red dashed line is the lower dispersion curve of molecules, obtained
via fitting each fixed-$k$ energy distribution curve (in \textbf{b})
with a two Gaussian distribution. It agree fairly well with the experimental
result (white symbols). \textbf{b}, Energy distribution curves for
selected values of $k$. \textbf{c}, The occupied density of state
(DOS). Blue dashed lines show the experimental peak positions.}

\label{fig2} 
\end{figure}

\begin{figure}[htp]
\begin{centering}
\includegraphics[clip,width=0.4\textwidth]{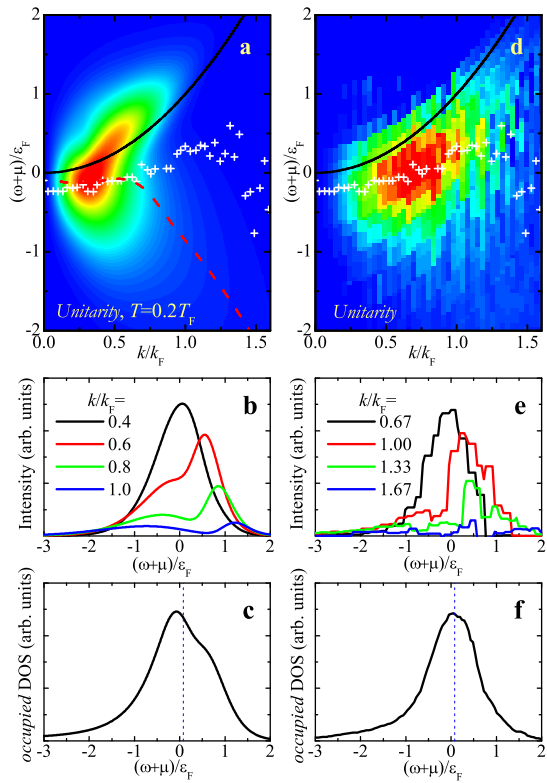} 
\par\end{centering}

\caption{(Color online) Single-particle excitation spectra of a strongly interacting
Fermi gas. \textbf{a-c}, Cluster expansion predictions ($z\simeq6$
and $\mu\simeq0.37\epsilon_{F}$). \textbf{d-e}, Corresponding experimental
data \cite{stewart}. In\textbf{ e}, for the experimental energy distribution
curves, we use a larger value of $k$ (i.e., enlarged by a factor
of 5/3) to account for a scaling discrepancy due to many-body correlations.}

\label{fig3} 
\end{figure}

For an \emph{absolute} comparison with experiment \cite{stewart},
we perform calculations using realistic experimental parameters, including
the measurement resolution. Fig. 2 presents the results on the BEC
side of crossover with $1/(k_{F}a_{s})=1.1$. The temperature $T=0.45T_{F}$
is estimated from an initial temperature $T_{i}=0.17T_{F}$ obtained
before the field sweep to the BEC side \cite{stewart}. The experimentally
observed upper and lower features, caused respectively by unpaired
atoms and molecules, are faithfully reproduced. In particular, the
experimental data for the quasiparticle dispersion of molecules, marked
by white symbols, agrees with our theory (lower red dashed line).
There is also a qualitative agreement for the energy distribution
curves (Figs. 2b and 2e) and the occupied density of states (Figs.
2c and 2f). A narrow peak due to free atoms and a broader feature
due to molecules are reproduced theoretically with very similar width
at nearly the same position. It is impressive that our simple quantum
cluster expansion is able to capture the main feature of the experimental
spectra. In contrast, a more complicated crossover theory with adjustable
parameters fails to account for the free-atom contribution with similar
parameters \cite{chen}.

Fig. 3 reports the spectra in the unitarity limit at the critical
temperature $T_{c}\simeq0.2T_{F}$. At such low temperatures, the
use of a cluster expansion becomes highly questionable as the fugacity
at the center $z\simeq6\gg1$. Nevertheless, we find that the dispersion
curve is lowered by the attractions by an amount comparable to the
Fermi energy $\epsilon_{F}$, as shown clearly by the red dashed line
in Fig. 3a. The calculated energy distribution curves bifurcates from
a single peak with increasing $k$ and becomes dominated by the lower
molecular branch (Fig. 3b), which eventually leads to the bending
back of the dispersion curve to negative energy. This picture is suggestive
of the existence of a pseudogap and is consistent with the experimental
findings (Fig. 3e). This surprisingly good agreement merits further
investigation. We conjecture that even at these relatively low temperatures
the virial expansion captures the dominant two-body correlations measured
in these experiments, apart from a possible overall scaling factor
due to the missing higher-order terms.

\begin{figure}[htp]
\begin{centering}
\includegraphics[clip,width=0.45\textwidth]{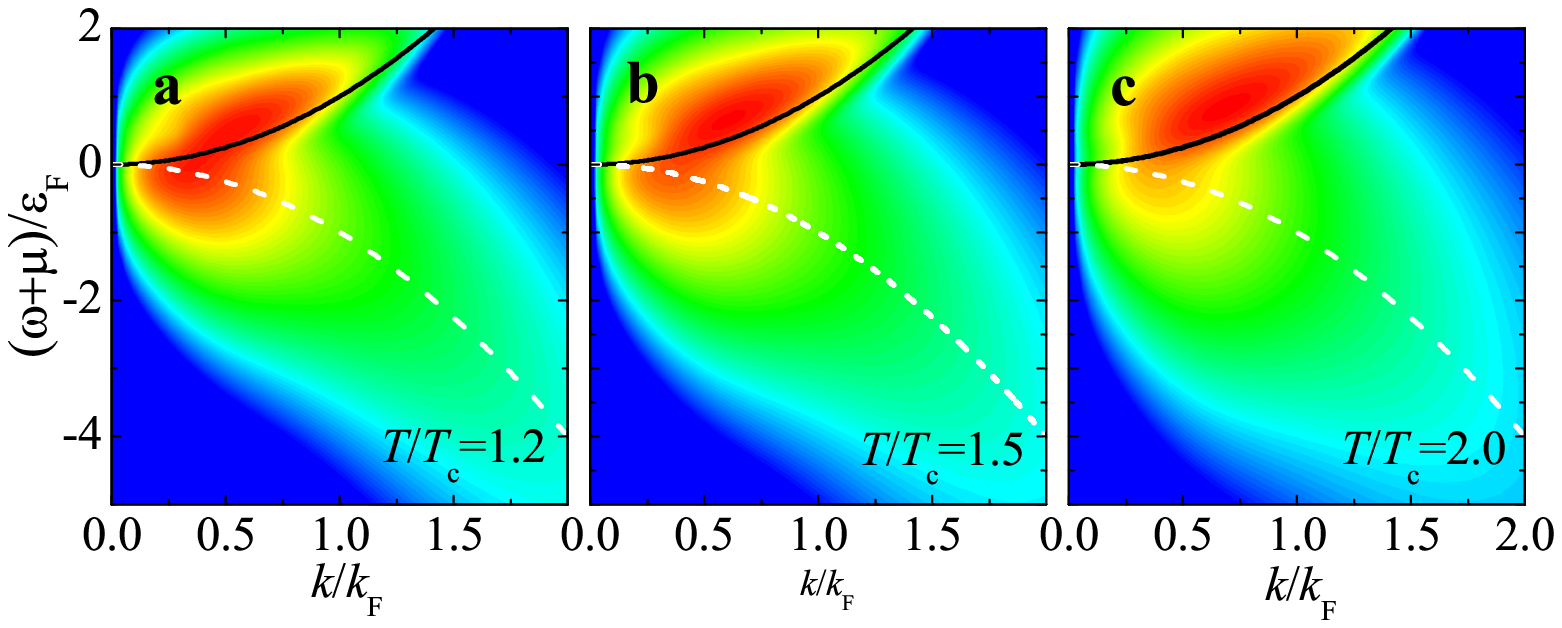} 
\par\end{centering}

\begin{centering}
\includegraphics[clip,width=0.45\textwidth]{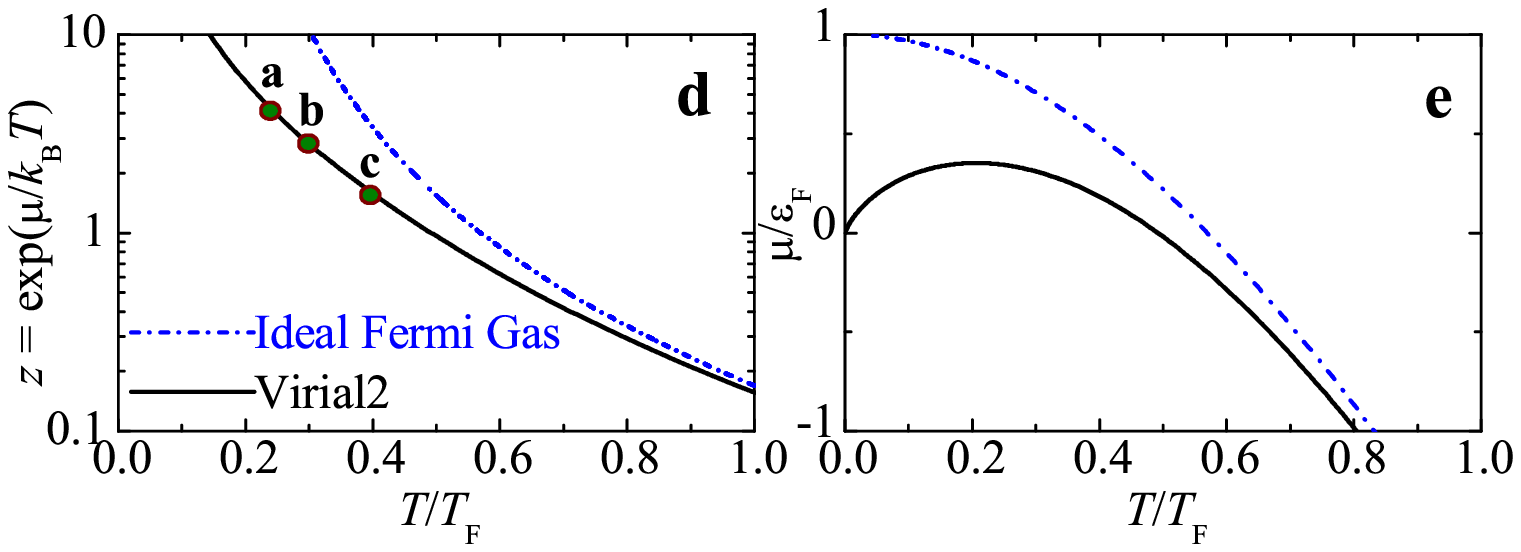}
\par\end{centering}

\caption{(Color online) Temperature dependence of the spectral intensity at
unitarity. \textbf{a-c}, The intensity increases from blue ($10^{-3}I_{max}$)
to red ($I_{max}$) in a logarithmic scale. \textbf{d} and \textbf{e},
Temperature dependence of the fugacity and chemical potential at unitarity.
Our leading cluster expansion appears to be applicable at $T\geq0.4T_{F}$,
where $z\sim1$ or $\mu\sim0$ \cite{hldnjp}. }

\label{fig4} 
\end{figure}

To study the temperature dependence of pseudogap pairing, we show
in Fig. 4 the unitary spectral intensity at $T/T_{c}=1.2$, $1.5$
and $2.0$, to be compared with the most recent measurement at JILA
\cite{gaebler}. In this logarithmic scale, a pseudogap structure
is clearly visible at about $k\sim0.5k_{F}$ and $\omega+\mu\sim-0.5\epsilon_{F}$.
The response is reduced with increasing temperature and becomes significantly
weak once $T>2T_{c}$. However, it seems difficult to accurately determine
a characteristic temperature above which the pseudogap response disappears.

In conclusion, the results obtained here provide a good qualitative
explanation for the recent single-particle spectral function measurement
in trapped strongly interacting ultracold fermions. It demonstrates
the precursor of fermionic pairing in the normal but strongly interacting
regime. In the near future, tomographic rf spectroscopy may be used
to reveal locally the homogeneous spectral function. Our calculation
can be extended to a uniform gas, for which we expect a stronger pairing
response and a wider temperature window for pseudogap. Our method
is also applicable directly to multi-component atomic Fermi gases
and can be used to understand the intriguing triplet and quadruplet
pairing in their dynamic responses. 

We gratefully acknowledge valuable discussions with P. Hannaford and
thank J. P. Gaebler and D. S. Jin for providing us with the experimental
data of Ref. \cite{stewart}. This work was supported by the ARC CoE
for Quantum-Atom Optics and ARC DP Nos. 0984522 and 0984637.

\textbf{Appendix. }\textit{General rules of cluster expansion}. ---
To calculate the Green function or correlation function, it is convenient
to separate out the contribution arising from interactions. To this
aim, for any physical quantity ${\cal Q}$ we may write ${\cal Q}=\{{\cal Q}\}^{(I)}+{\cal Q}^{(N)}$,
where the superscript {}``$N$'' in ${\cal Q}^{(N)}$ denotes the
part of a non-interacting system having the \emph{same} fugacity.
The operator $\{\}^{(I)}$ then picks up the residues due to interactions.
We then may write, \[
G\left({\bf r},{\bf r}^{\prime};\tau\right)=\left\{ G\left({\bf r},{\bf r}^{\prime};\tau\right)\right\} ^{\left(I\right)}+G^{\left(N\right)}\left({\bf r},{\bf r}^{\prime};\tau\right),\]
where $\{G\}^{(I)}$ can be expanded in terms of $\{X_{n}\}^{(I)}$.

For the single-particle spectral function, we determine first the
Green's function $G_{\sigma\sigma^{\prime}}({\bf r},{\bf r}^{\prime};\tau)$
by taking $\hat{R}=\hat{\Psi}_{\sigma}$ and $\hat{S}=\hat{\Psi}_{\sigma^{\prime}}$.
Then, we take the Fourier transformation with respect to $\tau$,
to obtain $G_{\sigma\sigma^{\prime}}({\bf r,r}^{\prime};i\omega_{n})$.
The spectral function is calculated via analytic continuation, \[
A_{\sigma\sigma^{\prime}}\left({\bf r},{\bf r}^{\prime};\omega\right)=-\frac{1}{\pi}\mathop{\rm Im}G_{\sigma\sigma^{\prime}}\left({\bf r},{\bf r}^{\prime};i\omega_{n}\rightarrow\omega+i0^{+}\right).\]
A final Fourier transform on ${\bf r}-{\bf r}^{\prime}$ leads to
$A_{\sigma\sigma^{\prime}}({\bf k},\omega)$, as measured experimentally.
For a normal, balanced Fermi gas, $A_{\uparrow\uparrow}=A_{\downarrow\downarrow}$
$\equiv A({\bf k},\omega)$ and $A_{\uparrow\downarrow}=0$.

\textit{Leading expansion of spectral function.} --- The leading term
of $\{G_{\uparrow\uparrow}({\bf r},{\bf r}^{\prime};\tau)\}^{(I)}$
takes the form, \[
-ze^{\mu\tau}\left\{ \text{Tr}_{1}\left[e^{-\beta{\cal H}}e^{\tau{\cal H}}\hat{\Psi}_{\uparrow}\left({\bf r}\right)e^{-\tau{\cal H}}\hat{\Psi}_{\uparrow}^{+}\left({\bf r}^{\prime}\right)\right]\right\} ^{\left(I\right)}.\]
 The trace has to be taken over all the single-particle states (i.e.,
$\psi_{p}$ with energy $\epsilon_{p}$) for a spin-down fermion.
We insert in the bracket an identity $\sum_{Q}\left|Q\right\rangle \left\langle Q\right|={\bf \hat{1}}$,
where $Q$ refers to the {}``paired'' state (i.e., $\Phi_{Q}$ with
energy $E_{Q}$) for two fermions with \emph{unlike} spins. It is
straightforward to show that, at the leading order, \[
\{G_{\uparrow\uparrow}\}^{(I)}=-ze^{\mu\tau}\sum_{p,Q}\left\{ e^{-\beta\epsilon_{p}+\tau\left(\epsilon_{p}-E_{Q}\right)}F_{pQ}\left({\bf r,r}^{\prime}\right)\right\} ^{\left(I\right)},\]
 where $F_{pQ}\equiv\int d{\bf r}_{1}d{\bf r}_{2}\psi_{p}^{*}({\bf r}_{1})\Phi_{Q}\left({\bf r},{\bf r}_{1}\right)\Phi_{Q}^{*}\left({\bf r}^{\prime},{\bf r}_{2}\right)\psi_{p}({\bf r}_{2})$.
Accordingly, the leading interaction correction to the spectral function,
$\{A\left({\bf k},\omega\right)\}^{(I)}$, is given by, \[
z\left(1+e^{-\beta\omega}\right)\sum_{p,Q}\left\{ \delta\left(\omega+\epsilon_{p}-E_{Q}+\mu\right)e^{-\beta\epsilon_{p}}\left|\tilde{F}_{pQ}\right|^{2}\right\} ^{(I)},\]
 where $\tilde{F}_{pQ}({\bf k})\equiv\int d{\bf r}d{\bf r}_{1}e^{-i{\bf k}\cdot{\bf r}}\psi_{p}^{*}({\bf r}_{1})\Phi_{Q}\left({\bf r},{\bf r}_{1}\right)$.

In an isotropic harmonic trap with frequency $\omega_{0}$, we solve
exactly the two-fermion problem for relative wavefunctions \cite{liuve}
and obtain $\{A\left({\bf k},\omega\right)\}^{(I)}$ using the above
procedure. In the end, we calculate \[
I({\bf k},\omega)=\frac{k^{2}}{2\pi^{2}}\left[\left\{ A\right\} ^{(I)}f\left(\omega\right)+A^{(N)}({\bf k},\omega)f\left(\omega\right)\right],\]
as measured experimentally \cite{stewart,gaebler}. Here, $f\left(\omega\right)=1/(e^{\beta\omega}+1)$
and the ideal spectral function $A^{(N)}=4\sqrt{2}\pi/(m^{3/2}\omega_{0}^{3})\left(\omega+\mu-\epsilon_{{\bf k}}\right)^{1/2}$.
To account for the experimental resolution, we further convolute $I({\bf k},\omega)$
with a Gaussian broadening curve.

In the BEC limit, we may show analytically that, \[
\left\{ A\right\} ^{(I)}f\propto\exp\left[-\beta\left(\sqrt{\epsilon_{{\bf k}}-\omega-\mu-\epsilon_{B}}-\sqrt{\epsilon_{{\bf k}}}\right)^{2}\right],\]
 where $\epsilon_{B}=\hbar^{2}/(ma_{s}^{2})$ is the binding energy.
Thus, at large $k$ the intensity peaks at $\omega+\mu=-\epsilon_{{\bf k}}-\epsilon_{B}$,
with a width $\sim\sqrt{k_{B}T\epsilon_{{\bf k}}}$.

\end{document}